# Pseudospin Polarization of Composite Fermions under Uniaxial Strain

Shuai Yuan,[1] Jiaojie Yan,[1,2] Ke Huang,[1,3] Zhimou Chen,[1] Haoran Fan,[1] L. N. Pfeiffer,[4] K. W. West,[4] Yang Liu,[1,*] Xi Lin[1,5,6,†]

[1]*International Center for Quantum Materials, Peking University, Beijing 100871, China*
[2]*Max Planck Institute for Solid State Research, Stuttgart 70569, Germany*
[3]*Department of Physics, The Pennsylvania State University, University Park, Pennsylvania 16802, USA*
[4]*Department of Electrical Engineering, Princeton University, Princeton, New Jersey 08544, USA*
[5]*CAS Center for Excellence in Topological Quantum Computation, University of Chinese Academy of Sciences, Beijing 100190, China*
[6]*Interdisciplinary Institute of Light-Element Quantum Materials and Research Center for Light-Element Advanced Materials, Peking University, Beijing 100871, China*

[*]liuyang02@pku.edu.cn
[†]xilin@pku.edu.cn

**Abstract**

*A two-dimensional system with extra degrees of freedom, such as spin and valley, is of great interest in the study of quantum phase transitions. The critical condition when a transition between different multicomponent fractional quantum Hall states appears is one of the very few junctions for many-body problems between theoretical calculations and experiments. In this work, we present that uniaxial strain induces pseudospin transitions of composite fermions in a two-dimensional hole gas. Determined from transport behavior, strain along <111> effectively changes pseudospin energy levels. We deduce that diagonal strain dominates these variations. Our experiment provides a wedge for manipulating two-dimensional interacting systems mechanically.*

In two-dimensional systems at ultra-low temperatures and high magnetic fields, exotic quantum states have been stabilized by the strong correlation between electrons. It is well known that non-interacting electrons form a series of discrete energy levels called Landau levels (LLs) in magnetic fields, and the filling factor *v* describes the number of occupied LLs. At different filling factors, two-dimensional electrons exhibit rich quantum phenomena, such as integer quantum Hall effect [1], fractional quantum Hall (FQH) effect [2] and charge density waves [3-8]. When a two-dimensional system has additional degrees of freedom, multicomponent quantum phases emerge. For example, at a particular filling factor, FQH states with different spin, valley, and subband compete with each other and transitions occur when those states become degenerate in energy. Such transitions can be induced by tuning the status of these components, such as tilted field [9-11], gate-tuning density [12, 13], biaxial strain [14, 15] and hydraulic pressure [16]. The energy separation between different possible states near the transition is universal, so its behavior can be compared with theoretical calculations, and provides an insight into a complex quantum system.

In general, holes in GaAs are subjected to strong spin-orbit interaction. The spin-orbit coupling in a two-dimensional system is affected by two kinds of asymmetry. One is bulk inversion asymmetry (BIA) which originates from the material's lattice structure and the other is structural inversion asymmetry (SIA) which depends on the effective out-of-plane electric field [17]. Uniaxial strain breaks the four-fold rotational symmetry and tunes the orbital motion of particles. Previous investigations on FQH states under strain mostly focus on symmetry, such as re-orientation of stripe phase [18] and anisotropy of Fermi contours [19-21]. However, the energy variation of FQH states under strain has been barely studied experimentally. The lowest two LLs share predominately $N = 0$ character and are separated by an energy separation $\Delta$ shown in Fig. 1(e). When $\Delta$ is small, multicomponent FQH effects appear. Here, we report pseudospin polarization transitions at $1 < \nu < 2$ induced by strain in a two-dimensional hole gas. Our result implies that the lattice deformation along $\langle 111 \rangle$, $\varepsilon_{\langle 111 \rangle}$, tunes $\Delta$ through the BIA and induces the observed transitions.

The GaAs/AlGaAs heterostructure sample in this work is grown on a GaAs (001) wafer by molecular beam epitaxy. Two-dimensional holes are confined to a 17.5-nm-wide symmetric GaAs quantum well, with carbon δ-doped layers and undoped $Al_{0.24}Ga_{0.76}As$ spacer layers on both sides. The carrier density is $n_p \simeq 1.6 \times 10^{11}$ cm$^{-2}$, and the mobility is larger than $1 \times 10^6$ cm$^2$V$^{-1}$s$^{-1}$ at low temperatures. Our rectangular sample is dissociated along [110] and [1$\bar{1}$0] which are denoted by *X*-axis and *Y*-axis. An L-shaped Hall bar is patterned in the $1 \times 1$ mm$^2$ center area of the sample, with 10 InZn contacts around the edge of the pattern. We use epoxy to bond the sample, as shown on the inset of Fig. 1(a). The sandwich structure is mounted on a commercial uniaxial strain cell (Razorbill CS130) with homemade thermal connections and electrical filters. The strain cell applies uniaxial tension (pressure) on the sample and makes the sample expand (shrink) along *X*-axis, while the sample along *Y*- and *Z*-axis is free and becomes shrunken (expanded). GaAs has a zinc-blende structure which consists of face-centered cubic lattices of Ga and As atoms. The As lattices are a translation of Ga lattices in the diagonal direction $\langle 111 \rangle$ of lattices. The strains along *X*-, *Y*-, *Z*-axis are projected to $\langle 111 \rangle$ to form the strain along $\langle 111 \rangle$, namely diagonal strain $\varepsilon_{\langle 111 \rangle}$ [22]. We note that the diagonal strain on the sample has a linear dependence on the deformation displacement in Fig. 1(c). We use $\varepsilon_{\langle 111 \rangle}$ to mark different strains applied on the sample in this measurement. The simulated $\varepsilon_{\langle 111 \rangle}$ along the two arms is nearly uniform, see Fig. 1(a).

Uniaxial strain breaks the four-fold rotational symmetry of the Fermi contour, which becomes elongated along the direction where tensile strain is applied [19-21]. Therefore, the longitudinal resistances versus strain at zero magnetic field are plotted in Fig. 1(b). When tensing the sample along *X*-axis, the resistance of *X*-arm increases rapidly as displacement $\Delta L_x$ increases, while the resistance of *Y*-arm decreases. The phenomenon is consistent with anisotropic Fermi contours under strain [19-21]. When the resistances of two arms are the same at $\Delta L_x = -0.68$ μm, the transport is isotropic suggesting zero uniaxial strain. The small negative shift of displacement results from the thermal expansion of the piezo, even though the strain cell uses the thermal-expansion-compensation technology [23].

The magnetoresistance traces of two arms at different $\varepsilon_{\langle 111 \rangle}$ for $1 < \nu < 2$ are shown in Fig. 2(a). All traces are measured at an effective temperature of about 41 mK using the same magnetic field ramping sequence. Transitions of FQH states around $\nu = 3/2$ can be seen. Corresponding $R_{xx}$ ($R_{yy}$) minima of FQH states disappear and reappear with varying $\varepsilon_{\langle 111\rangle}$. Note that the strength of the minima in the magnetoresistance traces is associated with the energy gaps of FQH states. Therefore, the energy gap undergoes a closing and re-opening. The weakening positions of FQH states in Fig. 2(a) are marked by black diamonds. From the $X$-arm data, the 5/3 minimum becomes less clear when $\varepsilon_{\langle 111\rangle}$ declines to $-5.4 \times 10^{-4}$, and it tends to disappear if $\varepsilon_{\langle 111\rangle}$ further decreases. On the other side of $\nu = 3/2$, the 4/3 minimum is distinct at low strains, but becomes less clear when $\varepsilon_{\langle 111\rangle}$ declines, disappears at $\varepsilon_{\langle 111\rangle} = -3.7 \times 10^{-4}$, and becomes distinct again at $\varepsilon_{\langle 111\rangle} = -5.4 \times 10^{-4}$. Other FQH states exhibit the same behaviors as the 4/3 state. The 8/5, 7/5, 11/7 and 10/7 minima disappear at $\varepsilon_{\langle 111\rangle} = -2.6 \times 10^{-4}$, $-1.5 \times 10^{-4}$, $-1.5 \times 10^{-4}$ and $-0.4 \times 10^{-4}$, respectively. These behaviors measured in $Y$-arm are slightly quantitatively different from that measured in $X$-arm, due to the insignificant different value of strain distribution shown in Fig. 1(a). Other than that, traces of $Y$-arm are qualitatively similar to those of $X$-arm.

The transitions in Fig. 2(a) data can be explained by the pseudospin polarization transitions of composite fermions (CFs). CF is a quasi-particle consisting of one electron and even magnetic flux quanta. The interacting electrons in a magnetic field correspond to non-interacting CFs in an effective magnetic field [24]. Similar to LLs formed by electrons, CFs form a series of discrete energy levels called $\Lambda$ levels and they are separated by CF's cyclotron energy $\hbar\omega_{CF}$. In Fig. 1(d), GaAs quantum well confines holes along the $Z$-axis and breaks the degeneracy of heavy-hole and light-hole bands. The two-dimensional holes occupy the doubly degenerate heavy-hole bands which consist of heavy heavy-hole (HHh) and light heavy-hole (LHh) subbands, see Fig. 1(d), and each subband generates a set of LLs. Within the axial approximation, the spin-orbit coupling mixes LLs with the same total angular momentum along the $Z$-axis, resulting in a complicated LL diagram, see Fig. 1(e). In particular, the two lowest LLs have mostly $N = 0$ characteristics and are separated by an energy $\Delta$ determined by the spin-orbit coupling strength. We denote them by up-pseudospin (↑) and down-pseudospin (↓) respectively. When $\Delta$ becomes comparable with $\Lambda$ level separation, multicomponent FQH states with pseudospin degree of freedom form [17].

Compared with LLs, the relation between composite fermion $\Lambda$ levels filling factor ($\nu^{CF}$) and electron LLs filling factor ($\nu$) is $\nu = \nu^{CF}/(2\nu^{CF} + 1)$ when $\nu < 1$. Since the particle-hole symmetry, $\nu$ equals $2 - \nu^{CF}/(2\nu^{CF} + 1)$ when $1 < \nu < 2$. The CFs also retain a pseudospin degree of freedom, and each $\Lambda$ level can be either up-pseudospin (↑) or down-pseudospin (↓). The $\nu = 5/3$ ($\nu^{CF} = 1$) state has only one $\Lambda$ level, and can be either up-pseudospin (↑) or down-pseudospin (↓) fully polarized, as shown in Fig. 2(b). Similarly, the $\nu = 4/3$ ($\nu^{CF} = -2$) state has three different pseudospin ground states, (↑↑), (↑↓), and (↓↓), and two pseudospin transitions.

In the CFs picture, the $\nu = 5/3$ state has two pseudospin ground states. Unlike other FQH states, the ground states of $\nu = 5/3$ are both fully polarized. The only transition between these two states, indicated by the increased longitudinal resistance, takes place when the energy separation of the pseudospin $\Delta$ equals zero. The transition from an unpolarized $\nu = $ 4/3 state (↑↓) to a fully polarized $\nu = 4/3$ state (↑↑ or ↓↓) is induced when the pseudospin energy separation $\Delta$ is equal to the $\Lambda$ level separation $\hbar\omega_{CF}$. Following the same rationale, we propose that $\Delta$ increases as varying $\varepsilon_{\langle 111\rangle}$ so that the 4/3, 8/5, 7/5, 10/7, and 11/7 states finally become fully polarized, illustrated as the energy diagram for CF $\Lambda$ levels in Fig. 2(c). We summarize the $R_{xx}$ normalized by resistance at $\nu = 3/2$ in Fig. 3(a). The maxima marked by the arrows in Fig. 3(a) correspond to the last transitions marked by circles in Fig. 2(c), after which the FQH states become fully polarized. From this figure, 5/3, 4/3, 8/5, 7/5, and 11/7 states are fully polarized at zero strain [22].

For further analysis, we summarize the strain at the transitions to fully polarized states and plot a phase diagram in Fig. 3(c). The diagram shows that critical $\varepsilon_{\langle 111\rangle}$ needed to polarize FQH states decreases as $|1/\nu^{CF}|$ increases, creating a "tent ($\Lambda$)" shaped phase boundary. FQH states except the $\nu = 5/3$ one are pseudospin-polarized above the tent, but pseudospin-partially-polarized below the tent. Du *et al.* found a similar spin transition induced by an in-plane magnetic field in a GaAs two-dimensional electron gas with consistent conclusions [9]. The scaled polarization energy $(\sqrt{n_p}/\sqrt{n_e})E_Z$ represents the energy separation $\Delta$ tuned by strain.

One might expect that $\Delta$ variation might be related to the anisotropic Fermi contour, where positive and negative in-plane strain $\varepsilon = \varepsilon_X - \varepsilon_Y$ correspond to anisotropy along the *X*- or *Y*-axis, respectively. The GaAs (001) surfaces have a square symmetry, so the *X*-axis is equivalent to the *Y*-axis geometrically. As shown in Fig. 1(b), the resistance anisotropy $|R_{xx} - R_{yy}|/|R_{xx} + R_{yy}|$ is an even function of the in-plane strain $\varepsilon = \varepsilon_X - \varepsilon_Y$ which is proportional to $\varepsilon_X$. Thus, if $\Delta$ were related to the anisotropy, one would expect it to be an even function of $\varepsilon_X$. However, the evolution in Fig. 3(a) and 3(b) and the phase diagram in Fig. 3(c) are clearly not evenly symmetric about $\varepsilon_X$. Therefore, the in-plane anisotropy is not likely the cause of variation in $\Delta$.

We suggest that the diagonal strain, $\varepsilon_{\langle 111\rangle}$ which monotonically depends on the displacement in Fig. 1(c), is likely the cause of the change of $\Delta$. In our nearly-symmetric GaAs quantum well sample, the effective SIA is negligible. The BIA caused by the built-in dipole electric field along $\langle 111\rangle$ direction in GaAs dominates the spin-orbit coupling. The energy separation can be written as

$$\Delta = \frac{1}{2}(E_z - \hbar\omega_c) + \frac{1}{2}\sqrt{(\hbar\omega_c + E_z)^2 + 8\tilde{\eta}^2 \frac{eB}{\hbar}}, \qquad (1)$$

where $\omega_c$ is cyclotron angular frequency, $E_z$ is Zeeman energy, and $\tilde{\eta}$ is the prefactor related to BIA [17]. We estimate that the cyclotron energy $\hbar\omega_c$ is about 1 meV and Zeeman energy $E_z$ is about -0.5 meV at $\nu = 1\sim 2$ in our sample. $\Delta$ equals $\sqrt{2\tilde{\eta}^2 eB/\hbar}$ at a small magnetic field, while it tends to $E_z$ at a high magnetic field. Since the effective $g$ factor of Zeeman energy is negative, $E_z$ is negative, so $\Delta = 0$ appears at some intermediate magnetic field, see Fig. 1(e). When we compress our sample along *X*-axis, the distance between Ga and As planes perpendicular to $\langle 111 \rangle$ direction decreases and the interatomic overlap increases. Our results suggest that the BIA-induced spin-orbit interaction $|\tilde{\eta}|$ increases and the energy degenerate point ($\Delta = 0$) moves to a higher magnetic field. The FQH states at $\nu = 1 \sim 2$ are on the high-field side of the degenerate point [25]. These states tend to be pseudospin-unpolarized when spin-orbit interaction increases. In a previous study, hydrostatic pressure can also reduce the pseudospin separation of two-dimensional hole systems [16]. In Fig. 4, we compare the evolution of $\nu = 5/3$ and $4/3$ states induced by uniaxial strain and the hydrostatic pressure. The excitation gap of $\nu = 5/3$ state decreases, and that of $\nu = 4/3$ state decreases to zero and then increases whether the sample is compressed uniaxially or uniformly. A comparable feature is the weakening of the $4/3$ state. It appears at 1.8 kbar in the hydrostatic pressure experiment, corresponding to critical $\varepsilon_{\langle 111 \rangle} = -7.9 \times 10^{-4}$, as well as at critical $\varepsilon_{\langle 111 \rangle} = -3.6 \times 10^{-4}$ in our uniaxial strain experiments [22]. We note that hydrostatic pressure makes the sample shrink uniformly, while uniaxial compressive pressure makes the plane perpendicular to $\langle 111 \rangle$ expand, which could be the result of the difference between the two experiments.

Examples with degrees of freedom controlled by strain are rare in two-dimensional systems. Electrons in the AlAs quantum well occupy two in-plane conduction-band valleys at the X point in the Brillouin zone, and strain is well known to induce transitions through the valley [15,26]. GaAs systems don't share the valley degree of freedom, but here we find strain-induced pseudospin transitions. Compared with other methods for tuning degrees of freedom, strain avoids lots of undesired effects. For example, the gate-tuning method varies both density and potential asymmetry together [12], and the tilted field method induces a finite-thickness effect to influence FQH states' stability [11]. Therefore, strain has the potential as a tool to probe multicomponent many-body states, such as non-Abelian even denominator states [27-30] and edge interactions between different polarization states [31].

In summary, we study the transport of two-dimensional hole gas systems confined in a symmetric quantum well. We observe the pseudospin transitions of CFs in a two-dimensional hole gas under strain, and propose that diagonal strain $\varepsilon_{\langle 111 \rangle}$ rather than the in-plane strain, plays a critical role in tuning the energy of different pseudospin bands. With the strain technique, precise mechanical control can be applied and we gain deeper insight into the complex energy structure of many-body states.

**Acknowledgments**

We thank P. Wang from Princeton University, A. Ward and J. Barraclough from Razorbill Instruments Ltd for discussions. We acknowledge Prof. R. Winkler's help and permission to use the valance band sketch and Landau level diagram. The work at PKU was supported by the National Key Research and Development Program of China (2021YFA1401900), the NSFC (12141001 and 11921005) and the Strategic Priority Research Program of Chinese Academy of Sciences (Grant No. XDB28000000). The work at Princeton University was funded by the Gordon and Betty Moore Foundation through the EPiQS initiative Grant GBMF4420, by the National Science Foundation MRSEC Grant DMR-1420541, and by the Keck Foundation.

**Figures**

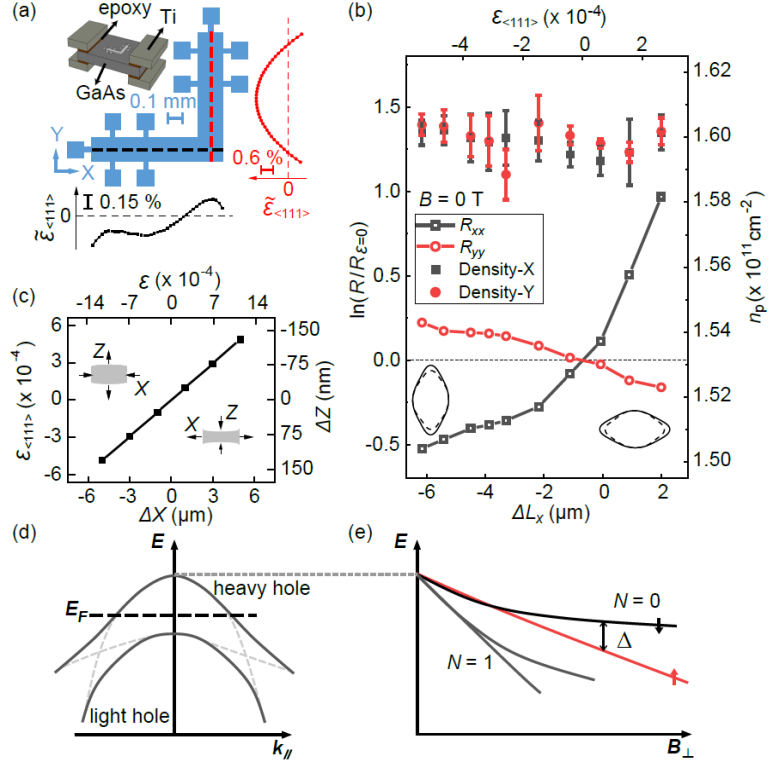

FIG. 1. Mechanical properties and transport properties of a GaAs sample. (a) Illustration of sample mounting, a pattern of an L-shaped bar, and strain distribution. Stress is applied along ⟨110⟩, paralleled with $X$-axis. When the sample is compressed, the distribution of diagonal strain $\tilde{\varepsilon}_{\langle 111 \rangle}$ along the black (red) dash line is plotted nearby. $\tilde{\varepsilon}_{\langle 111 \rangle}$ is defined as $\varepsilon_{\langle 111 \rangle}/\bar{\varepsilon}_{\langle 111 \rangle} - 1$. The diagonal strain $\varepsilon_{\langle 111 \rangle}$ is simulated through the software *Solidworks*, and $\bar{\varepsilon}_{\langle 111 \rangle}$ is the mean strain calculated from a $1 \times 1$ mm$^2$ area at the center of the sample. The configuration of the sample cell is shown as an inset. (b) Logarithm of longitudinal resistance $R_{xx}$ ($R_{yy}$) normalized by strain-free resistance $R_{\varepsilon=0}$, and hole density $n_p$, as a function of displacement $\Delta L_x$ at zero magnetic field. Traces of resistances cross at the zero-strain position. No apparent density shift is measured in our sample. Two small ellipses are diagrams of Fermi contour at positive and negative strain. The solid and dashed lines represent bands with different spins at $k = 0$. (c) Simulated mean diagonal strain, denoted by $\varepsilon_{\langle 111 \rangle}$, versus deformation displacement $\Delta X$ when applying uniaxial strain along $X$-axis. Simulated in-plane strain is defined as $\varepsilon = \varepsilon_X - \varepsilon_Y$. When compressing with a reasonable value of 5 μm, the sample is expected to have a significant $\varepsilon_{\langle 111 \rangle}$ up to $-4.9 \times 10^{-4}$. The insets show exaggerated deformation of the sample in the $X$-$Z$ plane. The difference between $\Delta L_x$ in (b) and $\Delta X$ in (c) suggests that the sample is subjected to a pre-pressurized status from the cooling down process. (d) Diagrams of valance bands in GaAs 2D hole systems. Since the quantization along $Z$-axis, the heavy-hole band and light-hole band show anticrossings and are mixed with each other [17]. (e) Landau level diagram of two-dimensional hole system. The nonlinear curves of Landau levels show strong spin-orbit coupling [17]. The arrows denote the pseudospins on the two lowest Landau levels. Δ is the energy separation between the two lowest Landau levels.

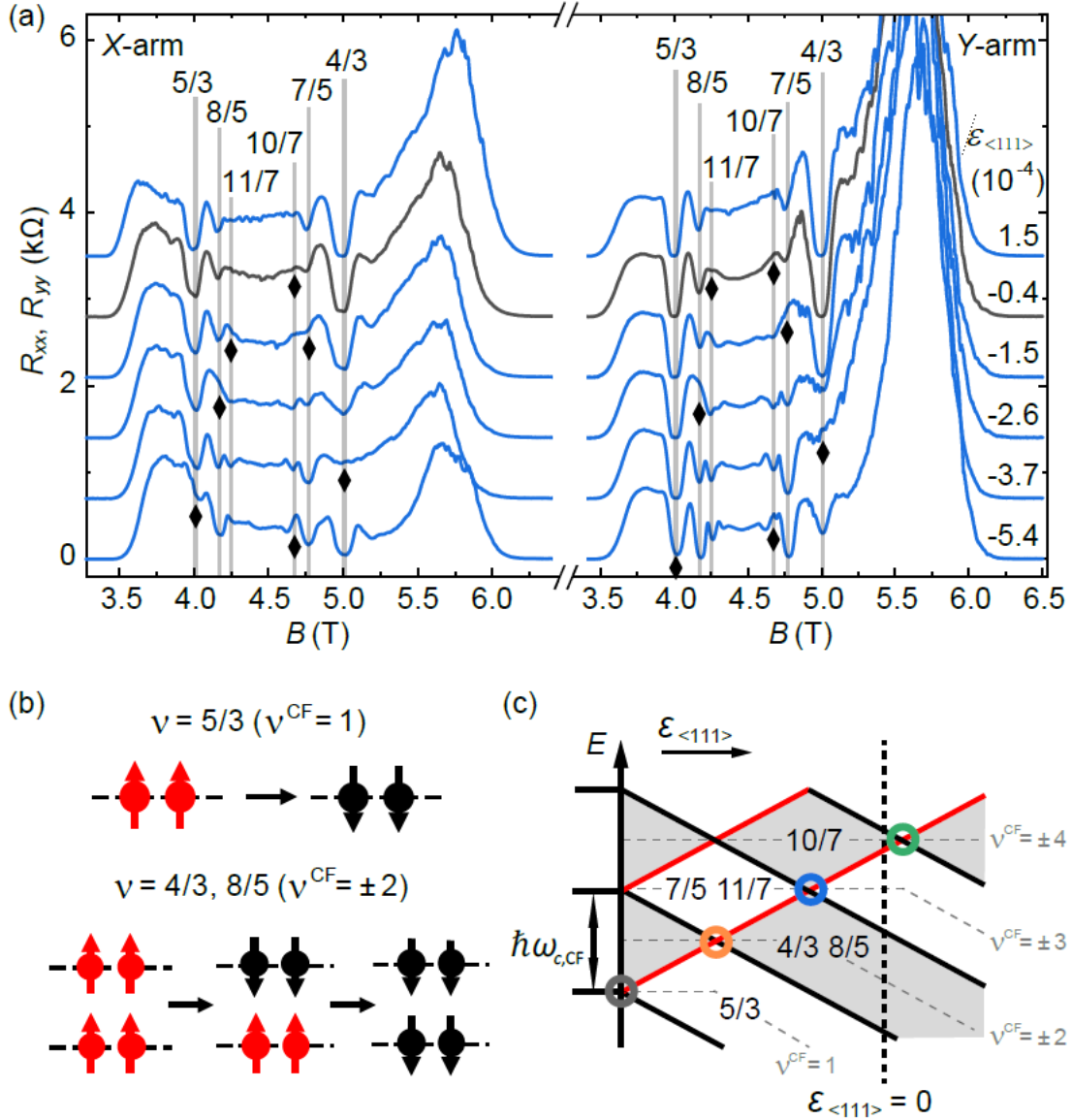

FIG. 2. (a) Longitudinal resistance traces of two arms measured under different strains at about 41 mK. Traces are offset vertically as multiples of 700 Ω for clarity. The values of applied diagonal strain $\varepsilon_{\langle 111\rangle}$ are shown on the right side of the Y-arm plot. The black trace shows data from the minimum strain. Possible FQH states are labeled by solid straight lines, based on the density calculated from Shubnikov–de Haas oscillations at low fields. Pseudospin transitions of FQH states take place at weakened minima of $R_{xx}$ and are marked with black diamonds. (b) Cartoon charts of CFs with pseudospin degree of freedom. (c) Energy diagram of CF's Λ levels. Black and red solid lines represent Λ levels with down- or up-pseudospin, respectively. Strain $\varepsilon_{\langle 111\rangle}$ induces pseudospin splitting of Λ level and pseudospin transition at the crossing.

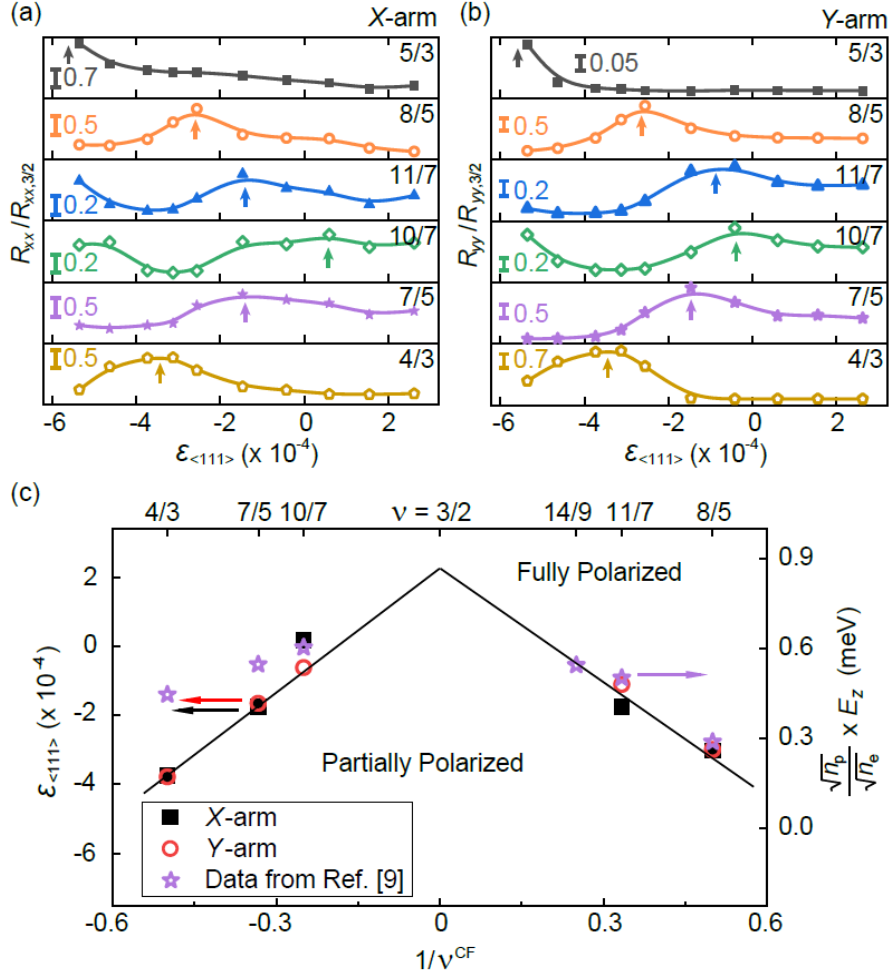

FIG. 3. (a) (b) Longitudinal resistance at different FQH states normalized to the nearly constant resistance of $\nu = 3/2$ as a function of strain $\varepsilon_{\langle111\rangle}$ along X-arm and Y-arm. The solid line in each trace is an eye guide for tendency. Arrows mark pseudospin transitions corresponding to circles in Fig. 2(c). (c) Phase diagram for the pseudospin polarization properties of CFs. The black square dots and red circular dots of the two arms are critical diagonal strains extracted from (a) and (b). The black solid lines drawn as an eye guide represent the phase boundary of the polarization transition. To compare data with spin transition in Ref. [9], the zero point of polarization energy is aligned with the minimum $\varepsilon_{\langle111\rangle}$ where pseudospin energy separation $\Delta \approx 0$. Moreover, we assume that polarization energy in two different systems is the same order of magnitude, so the polarization energy is scaled by the ratio of Coulomb energy which is proportional to the square root of carrier density, since Coulomb energy $V_c = e^2/(4\pi\epsilon l_B)$ and magnetic length $l_B = \sqrt{\hbar/eB} = \sqrt{\nu/2\pi n}$. The electron density in Ref. [9] is $n_e = 1.13 \times 10^{11}$ cm$^{-2}$, and the hole density in our sample is $n_p = 1.61 \times 10^{11}$ cm$^{-2}$.

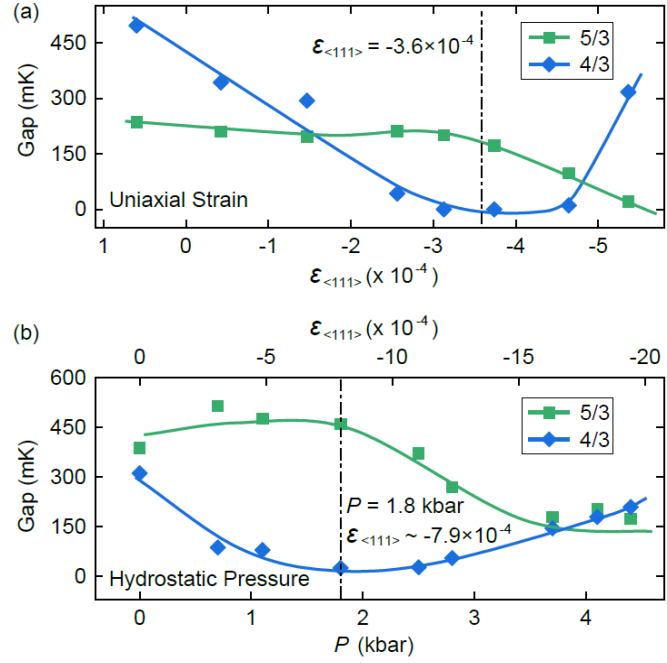

FIG. 4. Comparison of excitation gaps between uniaxial strain and hydrostatic pressure. (a) Excitation gaps of $\nu = 5/3$ and $4/3$ states under different uniaxial strains. The tendencies of traces are similar to those in hydrostatic pressure. (b) Data are from Ref. [16]. Since negative hydrostatic pressure, i.e. expanding samples uniformly, is hardly realized, the sample with hydrostatic pressure can only be compared with the compressed sample. The diagonal strains where the excitation gap of $\nu = 4/3$ state is minimum are comparable in the two experiments.

# Supplementary Information for

# "Pseudospin Polarization of Composite Fermions under Uniaxial Strain"

Shuai Yuan,[1] Jiaojie Yan,[1,2] Ke Huang,[1,3] Zhimou Chen,[1] Haoran Fan,[1] L. N. Pfeiffer,[4] K. W. West,[4] Yang Liu,[1,*] Xi Lin[1,5,6,†]

[1]*International Center for Quantum Materials, Peking University, Beijing 100871, China*
[2]*Max Planck Institute for Solid State Research, Stuttgart 70569, Germany*
[3]*Department of Physics, The Pennsylvania State University, University Park, Pennsylvania 16802, USA*
[4]*Department of Electrical Engineering, Princeton University, Princeton, New Jersey 08544, USA*
[5]*CAS Center for Excellence in Topological Quantum Computation, University of Chinese Academy of Sciences, Beijing 100190, China*
[6]*Interdisciplinary Institute of Light-Element Quantum Materials and Research Center for Light-Element Advanced Materials, Peking University, Beijing 100871, China*

[*]liuyang02@pku.edu.cn
[†]xilin@pku.edu.cn

Supplementary Information Contents

1. Simulation of sample mounting configuration
2. Tilted field experiments and spin-orbit coupling
3. Calculation on the strain along ⟨111⟩
4. Contrast of excitation gaps
5. Notes about theoretical interpretations of strained systems

1. Simulation of sample mounting configuration

The size of our GaAs two-dimensional hole gas (2DHG) sample is 3 mm × 2 mm × 0.5 mm. The sample is clamped at both ends by the titanium sample plates and glued with epoxy 2850. We use stainless steel screws to mount the whole structure on the strain cell. When the strain cell applies stress by changing voltage between piezoelectric stacks, the structure is stretched and the sample is subject to the strain. A parallel-plate capacitor is integrated into the strain cell and underneath the sample. The linearity and accuracy of capacitor have been checked through accurate calibration using a fiber-based interferometer before delivery. The capacitor could be used to measure the deformation displacement of the structure.

We use *Solidworks* to simulate the strain distribution on our sample. The model of the structure is shown in Fig. S1 and the materials' parameters of mechanical properties are displayed in Table S1. The simulated sample is a typical solid with similar mechanical properties as pure GaAs instead of an actual 2DHG sample. The simulated parameters are measured at room temperature. The Young's modulus of these materials increases slightly with decreasing temperature, so the simulation semi-quantitatively displays the strain distribution.

We set boundary conditions for the structure based on the working principle of the strain cell. We fix the left side of two left titanium plates and apply displacement $\Delta L_x$ on the right side of two right titanium plates. Typically, $\Delta L_x$ is about ± 5 μm, which is comparable to the maximum strain in our experiments. After meshing the structure, we calculate the strain value on each node. We extract the data from the node on the surface of the sample and cubically interpolate these 2D scattered data. The results are shown in Fig. S2. The 2DHG is confined in the L-shaped bar at the 1 mm² central area of the sample, where the strain homogeneity is less than 1.8%. We calculate the average strain at this area as the strain of the sample.

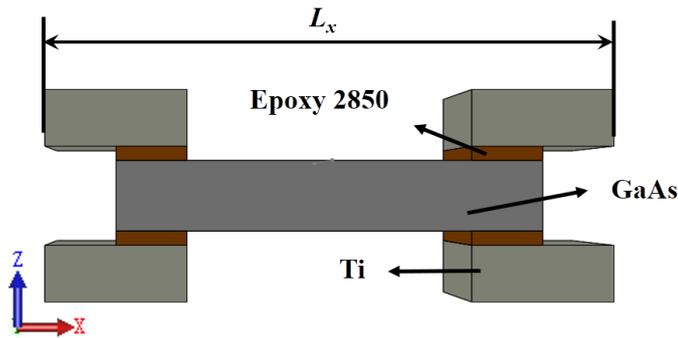

FIG. S1 Configuration of the sample mounting. The size ($X \times Y \times Z$) of Ti plate is 1 mm × 2 mm × 0.4 mm. The size of epoxy is 0.5 mm × 2 mm × 0.1 mm. The size of GaAs is 3 mm × 2 mm × 0.5 mm.

| Material | Elasticity Modulus (GPa) | Shear Modulus (GPa) | Poisson's Ratio | Density (g/cm³) |
|---|---|---|---|---|
| Ti | 105.0 | 39.0 | 0.33 | 4.51 |
| Epoxy 2850 | 15.0 | 6.0 | 0.30 | 2.91 |
| GaAs | 85.9 | 32.9 | 0.31 | 5.32 |

Table S1 Parameters of mechanic properties for simulation. The parameters of Ti are from *Solidworks* materials library. The parameters of epoxy 2850 are from Mark Edward Barber's thesis (Springer, 2018). The parameters of GaAs are from the following website *http://www.ioffe.ru/SVA/NSM/Semicond/GaAs/mechanic.html*.

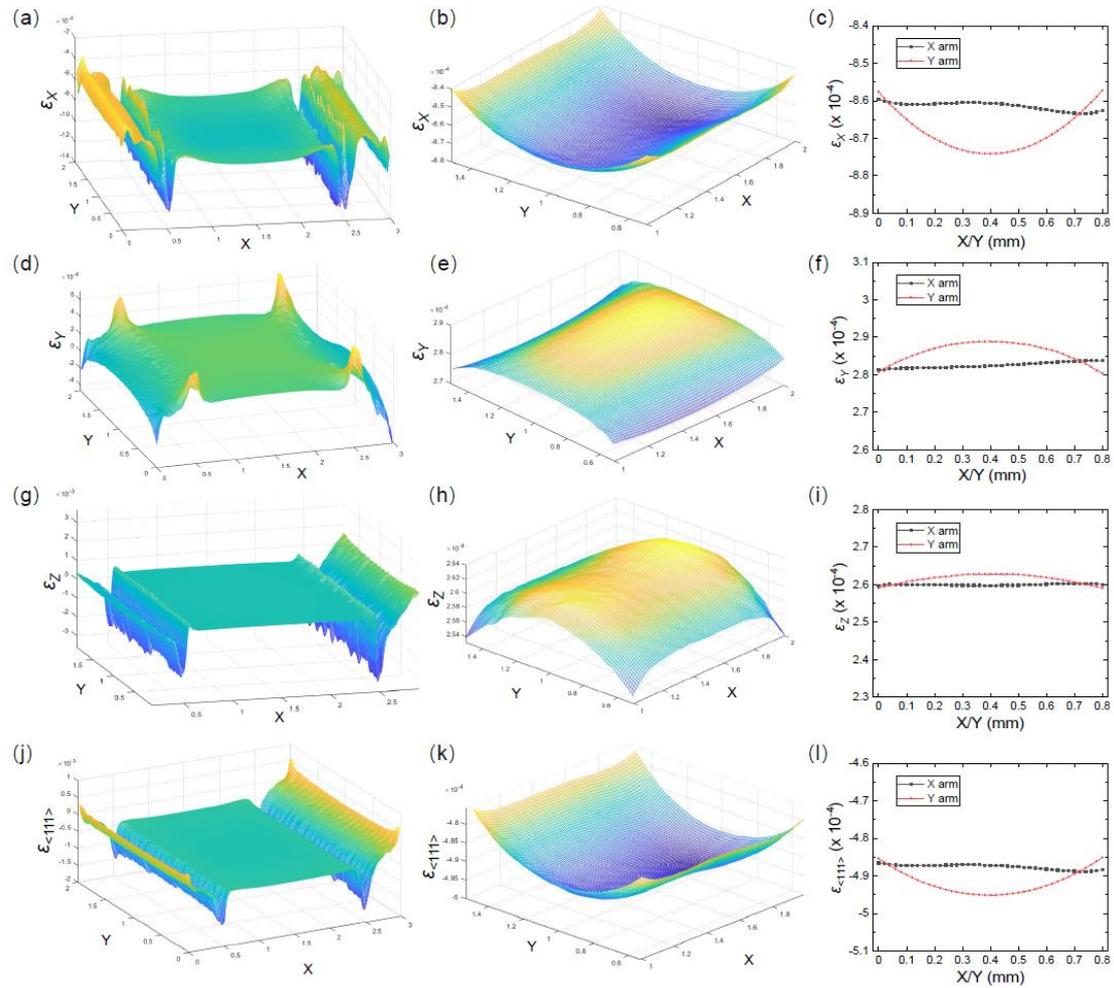

FIG. S2 Simulated results of strain distribution when we compress the sample along the $X$-axis with $\Delta L_x$ = -5 μm. (a) (d) (g) (j) show the distribution on the whole surface of GaAs. The areas of X = 0~0.5 mm and X = 2.5~3 mm are bonded with epoxy. (b) (e) (h) (k) show the distribution on the 1 mm² central area of the sample, i.e. the central data of (a) (d) (g) (j). (c) (f) (i) (l) are the strain distribution along the L-shaped bar corresponding to Fig. 1(a) of the main text. $\varepsilon_X$, $\varepsilon_Y$, and $\varepsilon_Z$ are the normal strains along $X$-, $Y$-, and $Z$-axis. $\varepsilon_{<111>}$ is calculated by the formula (3).

2. Tilted field experiments and spin-orbit coupling

In 2D electron systems, the Landau level spacing $\hbar\omega_c$ depends on the perpendicular component of the magnetic field $B_\perp$, but the Zeeman energy depends on the total magnetic field $B_{tot}$. Therefore, the two energy can be changed independently by tilting the sample. Keeping $B_\perp$ constant and increasing the tilted angle $\theta = \cos^{-1}(B_\perp/B_{tot})$ causes an increase in the total magnetic field $B_{tot}$, which increases the Zeeman energy [1].

With spin-orbit coupling, the Hamiltonian of our 2DHG sample is simplified as

$$\mathcal{H} = \hbar\omega_c \left(a^\dagger a + \frac{1}{2}\right)\begin{pmatrix}1 & 0 \\ 0 & 1\end{pmatrix} + E_z(B_\perp, B_\parallel) + \sqrt{\frac{2eB_{tot}}{\hbar}}\tilde{\eta}\begin{pmatrix}0 & a^\dagger \\ a & 0\end{pmatrix}, \quad (1)$$

where $\omega_c = eB_\perp/m^*$ is cyclotron angular frequency, $a^\dagger$ and $a$ are the ladder operators of Landau levels, and $\tilde{\eta}$ is the prefactor of the Dresselhaus effect term related to bulk-inversion asymmetry [2]. Because of strong spin-orbit coupling in 2DHG systems, Zeeman energy $E_z$ depends on in-plane and out-of-plane magnetic fields respectively. The energy of the Landau levels is

$$E_{N\pm} = \hbar\omega_c \left(N + \frac{1}{2} \mp \frac{1}{2}\right) \pm \frac{1}{2}\sqrt{(\hbar\omega_c + E_z)^2 + 8\tilde{\eta}^2\frac{eB_{tot}}{\hbar}\left(N + \frac{1}{2} \mp \frac{1}{2}\right)}, \quad (2)$$

where $N = 0,1,2,...$ is the index of Landau levels. For the $N = 0$ Landau level, in the limit of large magnetic fields, the pseudospin-splitting energy is $\Delta E = E_z$, which depends on the Zeeman effect. In the opposite limit $B_{tot} \to 0$, the energy is $\Delta E = 2|\tilde{\eta}|k_F$, which depends on the Dresselhaus term of spin-orbit coupling [2].

When we increase the tilted angle and the ratio $B_{tot}/B_\perp$, electrons on the low Landau levels tend to be polarized. If a ground state of a fractional quantum Hall (FQH) state is unpolarized at $\theta = 0°$, a small change in $B_{tot}/B_\perp$ will induce the resistance variation of FQH states, and it will become a fully polarized ground state at a large tilted angle where a transition will happen. But if a ground state is fully polarized, increasing the tilted angle will have little effect on the state. We also measure 2DHG sample in the tilted magnetic field, as shown in Fig. S3. The tilted angle is up to 48° where $B_{tot}/B_\perp$ increases by 50%, and the resistances of FQH states barely change. These phenomena imply the FQH states are likely polarized. It is noteworthy that the result of polarization is not definite, which needs more experiments. The result in the tilted field experiment is auxiliary and is not conflict with the strain experiments.

When we tune the spin-orbit coupling, polarization transitions also happen. Since $g^*$ is negative in GaAs, $\Delta E$ is positive in a zero field and negative in a high field. A zero of the pseudospin-splitting $\Delta E$ appears at some intermediate magnetic field [2]. Thus, when tuning the spin-orbit coupling, the energy of Landau levels with different pseudospins varies and degenerates for a specific magnetic field. If we compress the sample, the distance between atoms will become smaller, and the spin-orbit coupling will increase. Therefore, the deformation induced by compressive strain increases the $|\tilde{\eta}|$, and affects the pseudospin splitting of Landau levels.

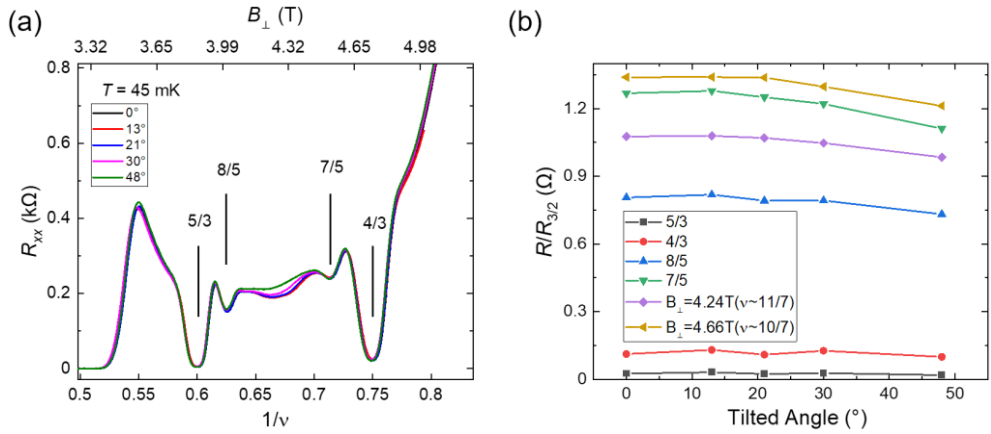

FIG. S3 (a) Longitudinal resistance at 45 mK under different tilted angles $\theta$ without strain. Traces with tilted angles are shifted vertically. $\theta$ is defined as the angle between the magnetic field and the out-of-plane direction of the sample. These data are from another sample on the same wafer. (b) Longitudinal resistance normalized to the resistance of $v = 3/2$ as a function of tilted angle. The data are extracted from (a).

3. Calculation on the strain along $\langle 111 \rangle$

Referring to the dissociated crystal orientation of our GaAs sample, we define that the $\langle 110 \rangle$ direction aligns with the *X*-axis, and the $\langle 1\bar{1}0 \rangle$ corresponds to the *Y*-axis. Based on the projective relation in space coordinates, the diagonal strain $\varepsilon_{\langle 111 \rangle}$ can be calculated as a function of normal strains. Ignoring higher-order terms of normal strains, we get

$$\varepsilon_{\langle 111 \rangle} = \sqrt{\frac{2}{3}(1+\varepsilon_X)^2 + \frac{1}{3}(1+\varepsilon_Z)^2} - 1 \approx \frac{1}{3}(2\varepsilon_X + \varepsilon_Z). \tag{3}$$

According to Fig. S2, $\varepsilon_X : \varepsilon_Y : \varepsilon_Z = 8.7 : -2.9 : -2.6$ under uniaxial strain in our experiments. The pseudospin transition of $\nu = 4/3$ states is at $\varepsilon_{\langle 111 \rangle} = -3.6 \times 10^{-4}$. Other strain components can be calculated by using the formula for coordinate transformation [4].

In hydrostatic pressure, $\varepsilon_X = \varepsilon_Y = \varepsilon_Z = \varepsilon_{\langle 111 \rangle}$. The $\nu = 4/3$ state under hydrostatic pressure disappears at 1.8 kbar [3]. The correlation formula [4] between hydrostatic pressure $p$ and uniaxial strain $\varepsilon$ is

$$\frac{p}{\kappa} = \frac{\Delta V}{V} = 1 - (1+\varepsilon_X)(1+\varepsilon_Y)(1+\varepsilon_Z) \approx -3\varepsilon_{\langle 111 \rangle} \tag{4}$$

$\kappa$ = 75.5 GPa is GaAs bulk modulus. Therefore, strain along $\langle 111 \rangle$ is equal to $-7.9 \times 10^{-4}$ when the $\nu = 4/3$ state becomes weaker.

4. Contrast of excitation gaps

By measuring the $R_{xx}$ minimum of FQH states as a function of temperature, the excitation gaps are derived from the Arrhenius function $R_{xx} \propto \exp(-E_{Gap}/2k_BT)$. We measure the excitation gaps of FQH states in $v = 1\sim2$ under different strains, as shown in Fig. S4(a). The $v = 5/3$ and 4/3 states are strong near zero strain, while other states are weak. With the decrease of diagonal strain, 5/3 and 4/3 states become weaker, while 7/5 and 8/5 states become stronger. It's worth noting that the 4/3 state disappears at $\varepsilon_{\langle 111 \rangle} = -3.6 \times 10^{-4}$ and reappears at lower diagonal strain. These phenomena are similar to FQH states under hydrostatic pressure $P$ = 0~4 kbar, as shown in Fig. S4(b). Compressing the sample uniaxially is consistent with compressing the sample uniformly, which both can degenerate the pseudospin. Hydrostatic pressure changes the confinement potential along the out-of-plane direction, and tunes the warping of energy bands in the in-plane Brillouin zone by deforming the lattice structure at the same time. These complex effects mix and make us hard to distinguish the mechanism between in-plane and out-of-plane actions. Uniaxial strain breaks the symmetry and helps us exclude the possibility of in-plane strain, so that we deduce that the diagonal strain dominates the pseudospin transition. Moreover, combining the data in Fig. 3 of the main text, we find more pseudospin transitions under the uniaxial strain.

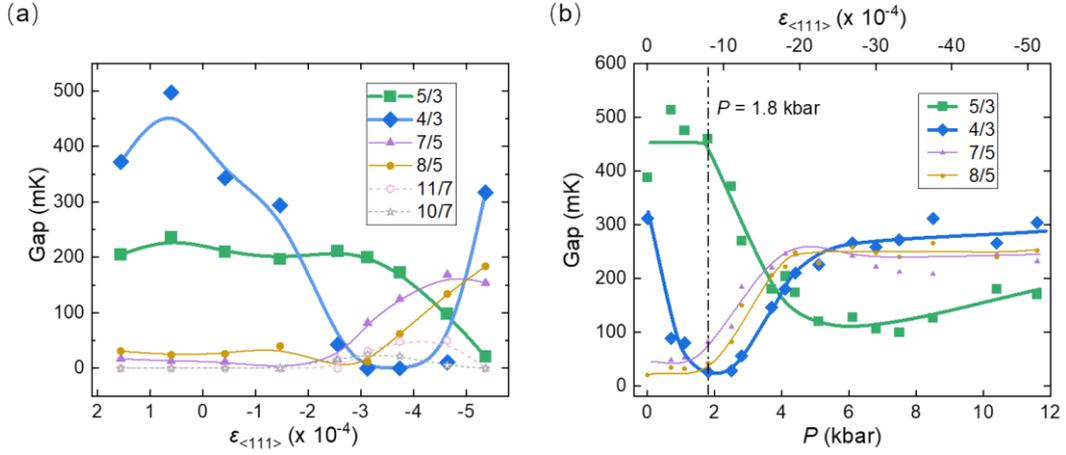

FIG. S4 Excitation gap of FQH states under (a) uniaxial stress and (b) hydrostatic pressure. The data of (b) is from Ref. [3].

5. Notes about theoretical interpretations of strained systems

Pseudospin transitions happen when strain is applied in our experiments. Before analyzing the strain effects, it is necessary to comprehend the characteristics of Landau levels in 2DHG systems. The topmost valence band is the heavy-hole (HH) band shown in Fig. 1 of the main text. At the $\Gamma$ point and $B = 0$ T, the pure HH bands corresponding to spinor components $s = \pm 3/2$ and pure light-hole (LH) bands with $s = \pm 1/2$ are considered as eigenstates. However, it is notable that the FQH states observed in high magnetic fields may not solely originate from the Landau levels of pure HH bands. The eigenstates of Hamiltonian in the magnetic fields are a linear combination of HH and LH states, which can be expressed as:

$$\Psi_{\mathcal{N}} = \sum_s | N = \mathcal{N} - s - \frac{3}{2} \rangle \xi_s^{\mathcal{N}} u_s, \qquad (5)$$

where $|N\rangle = 0$ when $N < 0$. $\mathcal{N} = 0, 1, 2 \ldots$ is the quantum number of LLs, $\xi_s^{\mathcal{N}}$ is the weight of HH and LH spinors and $u_s$ is the band-edge Bloch function. Due to spin-orbit interaction, the Landau levels spawned from different spinors become mixed through the coefficient $\xi_s^{\mathcal{N}}$ [5]. The various $|N = 0\rangle$ states are the result from the mixing of four states $|\mathcal{N}, s\rangle = |3, 3/2\rangle, |2, 1/2\rangle, |1, -1/2\rangle, |0, -3/2\rangle$ with different $\xi_s^{\mathcal{N}}$. Therefore, they are considered as multicomponent phases of the $N = 0$ Landau levels. This is why we use pseudospins instead of spins to describe these distinct phases. The observed transitions in our experiments arise from the energy competitions among these multicomponent phases under strain.

The classical band theory in semiconductors under strain is the Bir-Pikus theory [6]. The Bir-Pikus theory calculates the strained band structure in the **k·p** framework. It can effectively describe the shifts and warping of bands under strain, encompassing both HH and LH bands in three-dimensional semiconductors. However, this theory can't adequately interpret our data for the following reasons.

(1) We specifically focus on investigating the two spin-splitting bands of HH and their Landau levels. This splitting arises not only from the presence of a magnetic field but also due to the influence of spin-orbit coupling in 2D systems [2]. In Bir-Pikus theory, however, strain doesn't lift spin-degeneracy of HH and LH, and the two bands remain doublets around the $\Gamma$ point. The theory can't explain the splitting of HH band under strain.

(2) In Bir-Pikus theory, the common factor between hydrostatic pressure and uniaxial strain is that both alter the volume of the sample, and add a deformation potential on the valence band edge. The energy variation is proportional to the trace of the strain tensor, $\Delta E_\varepsilon = a_v(\varepsilon_{xx} + \varepsilon_{yy} + \varepsilon_{zz})$. The same phenomenon, weakened and then strengthened FQH states, is observed in both our uniaxial strain experiments and the hydrostatic pressure experiments of Huang et al. [3]. Referred to the pseudospin transition of 4/3 state, the trace of strain tensor can be easily calculated according to Supplementary Information 3. The critical trace equals $\mathrm{tr}(\varepsilon) = \varepsilon_{xx} + \varepsilon_{yy} + \varepsilon_{zz} = \varepsilon_X + \varepsilon_Y + \varepsilon_Z \approx -2.4 \times 10^{-3}$ under hydrostatic pressure, while it is

tr($\varepsilon$) ≈ $-2.3 \times 10^{-4}$ under uniaxial strain. They differ by an order of magnitude but exhibit the same phenomenon, which the theory can't explain.

(3) Bir-Pikus theory doesn't provide any information on Landau levels formed by valance bands. As emphasized above, the Landau levels are not the eigenstates of Hamiltonian in the zero field, but the mixing of different bands. The band theory at $B$ = 0T doesn't adequately explain our experiments.

Although we are unable to perform calculations for Landau levels under strain due to computational challenges, we provide a consistent interpretation in the main text. The spin-orbit coupling in GaAs accounts for the energy variation of Landau levels with different pseudospins. It is closely associated with bulk-inversion-asymmetry, and the diagonal strain (strain along ⟨111⟩) modulates the strength of this asymmetry. The diagonal strain is comparable in both uniaxial strain experiments and hydrostatic pressure experiments. To validate the strain effect, accurate self-consistent calculations of Landau levels from theorists are anticipated.